\documentclass[twocolumn,showpacs,preprintnumbers,amsmath,amssymb]{revtex4}

\usepackage{graphicx}
\usepackage{dcolumn}
\usepackage{bm}

\begin{document}


\title{Enhancement of surface activity in CO oxidation on
Pt(110)\\through spatiotemporal laser actuation}

\author{L. Qiao}
\author{X. Li}
\altaffiliation[Present Address: ]{Merck Research Laboratories,
West Point, Pennsylvania, PA 19486, USA}
\author{I. G. Kevrekidis}
\altaffiliation[Also at ]{the Program in Applied and Computational
Mathematics(PACM) , Princeton University, Princeton, NJ 08544,
USA}
\author{C. Punckt}
\affiliation{Department of Chemical Engineering, Princeton
University, Princeton, NJ 08544 USA}
\author{H. H. Rotermund}
\affiliation{Department of Physics and Atmospheric Science,
Dalhousie University Halifax, Nova Scotia, B3H 3J5 Canada}

\date{\today}

\begin{abstract}
We explore the effect of spatiotemporally varying substrate temperature
profiles on the dynamics and resulting reaction rate enhancement
for the catalytic oxidation of CO on Pt(110).
The catalytic surface is ``addressed" by
a focused laser beam whose motion is computer-controlled.
The averaged reaction rate is observed to undergo a characteristic
maximum as a function of the speed of this moving laser spot.
Experiments as well as modelling are used to explore and rationalize
the existence of such an optimal laser speed.

\end{abstract}

\maketitle

\section{Introduction}

A crucial component of chemical engineering process design is to locate
optimal operating conditions, which yield optimized productivity,
selectivity and flexibility under safety and environmental
constraints.
Typically one searches for optimal {\em steady state} conditions,
possibly stabilized through feedback loops;
yet it is also well-known that non-steady-state operating
conditions may lead to better average
performance~\cite{Bailey73}.
Examples of such chemical engineering processes include fast
pressure swing adsorption~\cite{Ruthven93} and reverse flow
reactors~\cite{Matros96}.
Improvement of the average process performance under
non-steady-state conditions (e.g. periodically varying reaction
conditions) can sometimes be rationalized through certain
resonances between the characteristic operation time scale (e.g.
the period of forcing) and the intrinsic time scales of the system
\cite{Kevrekidis86}.

Catalytic reaction systems often exhibit rich nonlinear dynamic
behavior, including spatiotemporal patterns such as
propagating reacting fronts, pulses, rotating spirals and even
chaos~\cite{Cross,Imbihl,Imbihl95}.
Since the emergence of these spatiotemporal patterns is governed
by the system intrinsic time and length scales, catalytic systems
may be promising testing fields for optimal non-steady-state
operating policies~\cite{Machado05}.
The development of advanced surface resolving
techniques such as PEEM~\cite{peem,Engel91}, EMSI~\cite{RAM}, and
RAM~\cite{Punckt07} has facilitated the observation of
catalytic spatiotemporal patterns at the micron scale.
Related advances in spatiotemporally addressing catalytic
surfaces at the micron scale, in particular through  computer-controlled laser beams,
have enabled us to spatiotemporally modify local reaction conditions
in real time: pulses and fronts, the building blocks of
spatiotemporal patterns, can be initiated, erased or guided
through appropriate local laser actuation in space and
time~\cite{Science2001}.

In a previous experimental study of modifying the local catalytic surface
activity in CO oxidation on Pt(110) through local laser actuation,
it was discovered that optimal laser scanning policies exist,
maximizing overall reaction (i.e. CO$_2$ production) rate when
the system is close to the excitable regime~\cite{twists}.
The reaction conditions used in those experiments were selected in a way
such that the system was initially at a CO-poisoned state (with Pt surface
dominated by adsorbed CO which leads to low CO$_2$ production
rate), but could become locally excitable upon local laser
heating.
The actuation of the laser beam converts the local Pt
surface to an excitable state by increasing the local temperature
and hence accelerating the desorption of CO.
When the laser beam is kept at a fixed position, reacting oxygen pulses
are periodically initiated around the laser heating center, and
propagate away with significant higher local reaction rate.
When the laser beam is scanned at a constant speed, it continuously
locally excites the Pt surface ~\cite{dragging}.
Depending on the intensity of the laser,
the spatiotemporal ``policy"  at which it scans the surface,
its scanning speed, and the characteristic times
required for the surface to return -after excitation- to the
quenched state, different maxima in the overall CO$_2$
production rate were observed and partially rationalized in Ref.~\cite{twists}.
In order to systematically explore optimal operating
conditions we choose a simple laser scanning policy
(constant speed ``dragging" along a one-dimensional path, ~\cite{dragging})
and attempt to establish
a quantitative connection between the laser scanning
policy and the corresponding overall CO$_2$ production rate.
This is attempted through modelling of the appropriate catalytic
reaction-diffusion system with spatiotemporal laser actuation.
As a first step, we studied the simple case in which the laser was
scanned in a fixed circular pattern at various speeds, through
both numerical bifurcation analysis and simulations.
By identifying a certain pulse instability associated with the
dragging of the laser spot, our computational results could
predict the existence of a local maximum in overall CO$_2$
production rate while varying the laser scanning speed,
in agreement with the experiments.
The paper is organized as follows.
We begin with a brief description of our model of CO oxidation on
Pt(110) with local laser dragging.
Computational results from numerical bifurcation analysis and
transient simulations are then presented and discussed.
We also briefly describe the experimental setup and present the
experimental results; we conclude with a discussion of the relation
between experiments and modeling, and possible extensions of the work.

\section{Modelling}

The experimental system of interest is the UHV oxidation of CO on
Pt(110); we therefore use the Krischer-Eiswirth-Ertl (KEE)
reaction-diffusion model for this reaction \cite{KEE_model}.
The surface reaction follows a Langmuir-Hinshelwood mechanism:
\begin{eqnarray*}
{\rm CO}+*&\leftrightharpoons& {\rm CO}_{ads},\\
2*+{\rm O}_2&\rightarrow& 2{\rm O}_{ads},\\
{\rm CO}_{ads}+{\rm O}_{ads}&\rightarrow& 2*+{\rm CO}_2\uparrow,\\
\end{eqnarray*}
accompanied by a $1\times2\rightarrow1\times1$ phase transition of
the Pt(110) surface due to CO adsorption.
The equations for this kinetic model are
\begin{equation}
\dot{u}=k_us_up_{\rm CO}\left[1-\left(\frac{u}{u_s}\right)^3\right]-k_1u-k_2uv+\nabla\cdot(D_u\nabla
u),\label{eqn_u}\end{equation}
\begin{equation}
\dot{v}=k_vp_{\rm O_2}[ws_{v_1}+(1-w)s_{v_2}]\left(1-\frac{u}{u_s}-\frac{v}{v_s}\right)^2-k_2uv,\label{eqn_v}
\end{equation}
\begin{equation}
\dot{w}=k_3(f(u)-w),\label{eqn_w}
\end{equation}
where $u$, $v$ and $w$ denote the surface coverage of CO, O,
and the surface fraction of the $1\times1$ phase respectively.
The adsorption rate constants for CO and O$_2$, $k_u$ and $k_v$
respectively, are set to fixed values within the temperature range
considered in this paper.
The rate constants $k_1,k_2$ and $k_3$
for the desorption, reaction and surface phase transition are
temperature dependent through the Arrhenius formula $k_i=k^0_i\exp(-E_i/kT)$;
$T$ is
the temperature of the single crystal.
We used the parameters for Pt(110) given in Table II of
Ref.~\cite{KEE_model} as follows:\\

\begin{center}
\begin{tabular}{l}
$u_s=1$, $v_s=0.8$, $s_u=1$, $s_{v_1}=0.6$, $s_{v_2}=0.4$,\\\\
$k_u=3.135\times10^5$ s$^{-1}$ mbar$^{-1}$,\\\\
$k_v=5.858\times10^5$ s$^{-1}$ mbar$^{-1}$,\\\\
$k_i=k^0_i\exp(-E_i/kT)$, $i=1$, 2, and 3,\\\\
$k^0_1=2\times10^{16}$ s$^{-1}$, $E_1=38$ kcal/mol,\\\\
$k^0_2=3\times10^{6}$ s$^{-1}$, $E_2=10$ kcal/mol,\\\\
$k^0_3=10^2$ s$^{-1}$, $E_3=7$ kcal/mol.~~~~~~~~~~~~~~~~~~~~~~~~~\\\\
\end{tabular}
\end{center}
We adopted the diffusion coefficients reported in
Ref.~\cite{Oertzen94}:
\[D_u=D^0_u\exp(-E_u/kT)\]
where, in the $[1\bar10]$ direction, $D^0_u=5\times10^{-3}$ cm$^2$/s,
$E_u=10$ kcal/mol and in the $[001]$ direction $D^0_u=7\times10^{-4}$
cm$^2$/s, $E_u=8.9$ kcal/mol.
The function $f(u)$ has been fitted to experimental data to give
the rate of surface phase transition as a function of $u$, the
coverage of CO, as follows:

\[f(u)=\left\{\begin{array}{ccc}0&\mbox{for}&u\leqslant0.2,\\
\frac{u^3-1.05u^2+0.3u-0.026}{-0.0135}&\mbox{for}&0.2<u<0.5,\\
1&\mbox{for}&u\geqslant0.5.\end{array}\right.
\]

In Eqs.~(\ref{eqn_u})-(\ref{eqn_w}) the temperature field is
assumed to be homogeneous across the catalytic surface and
prescribed by a single parameter $T$.
When a laser beam is scanned across the sample surface, the
local temperature of the catalytic surface around the laser spot is
temporarily increased, leading to spatiotemporal heterogeneity in the surface
temperature field.
In order to model such a spatiotemporal heterogenous temperature
field, we make the following two simplifications
\cite{Cisternas03A,Cisternas03B}:
a) the heat generated by the CO oxidation reaction on the Pt(110)
surface ($\sim 1$ mW/cm$^2$) is neglected compared to the power of
the laser beam ($\sim 1$ W for a crystal surface of 75 $\mu$m in
diameter (size of the laser spot));
b) we consider that the local temperature profile inhomogeneity is established
(and then vanishes) instantly as the laser beam is applied (and then removed).
This is because the time scale associated with the local
temperature change ($\sim 10^{-3}$ second, estimated from
$W^2/\alpha$, where $W$ is the diameter of the laser spot (75
$\mu$m) and $\alpha$ is the thermal diffusivity of Pt ($\sim
10^{-1}$ cm$^2$/s)) is much smaller than the time scales of local
adsorbate coverage change ($\sim 10^{-1}$ second and $\sim 1$
second, estimated from CO desorption with 1/$k_1$ and CO diffusion
with $W^2/D_{u}$.
%
%
%
Based on these two simplifications, the spatiotemporal change in
the temperature field induced by a moving laser beam can be
modelled as a temperature ``bump" with fixed shape co-travelling
with the laser spot.

In the experiments, the laser beam is moved in a circular path on
the sample surface at constant speed.
Because the typical diameter of the circular path ($\sim$ 860 $\mu$m) is
significantly larger than the size of the laser spot ($\sim$ 75 $\mu$m in
diameter), it is informative to model the laser scanning in one dimension
with periodic boundary conditions.
A computational frame co-travelling with the laser spot is introduced.
Based on the KEE model and the discussion above, we have the
following equations for CO oxidation in a one-dimensional periodic
domain with laser scanning in a co-moving frame:

\begin{equation}
u_t=k_us_up_{\rm
CO}\left[1-\left(\frac{u}{u_s}\right)^3\right]-k_1u-k_2uv+(D_uu_z)_z+
cu_z,\label{cotravel_u}\end{equation}
\begin{equation}
v_t=k_vp_{\rm
O_2}[ws_{v_1}+(1-w)s_{v_2}]\left(1-\frac{u}{u_s}-\frac{v}{v_s}\right)^2-k_2uv+cv_z,\label{cotravel_v}
\end{equation}
\begin{equation}
w_t=k_3(f(u)-w)+cw_z,\label{cotravel_w}
\end{equation}
where $c$ is the speed of the co-moving frame (equal to the laser
moving speed); $z$ is the spatial coordinate in the co-moving
frame with $z\equiv x-ct$; and the remaining symbols are the same as in
Eqs.~(\ref{eqn_u})-(\ref{eqn_w}), except $k_1$, $k_2$ and $k_3$,
which are now spatially dependent because of the nonuniform
temperature field $T(z)$ due to local laser heating.
In the computations the local $T(z)$ heterogeneity due to the
laser spot is approximated by a Gaussian profile.

\section{Computational results}

\begin{figure}
\centering
\includegraphics[width=0.9\columnwidth]{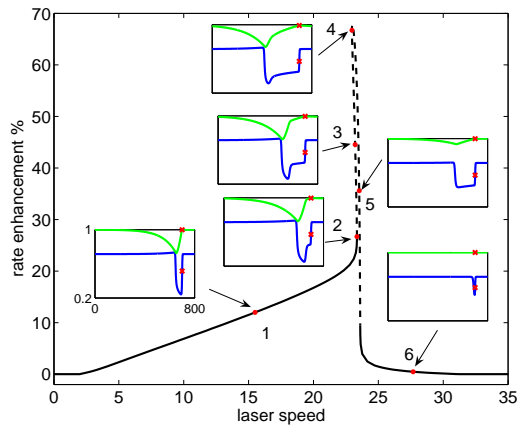}
\caption{(Color online) Enhancement in the CO$_2$ production rate as a function of
the laser scanning speed for isotropic CO diffusion.
The steady coverage profile is computed in a frame
co-travelling with the laser spot.
Stable (unstable) steady coverage profiles are marked in solid (dashed) lines.
Representative coverage profiles are shown in the insets, where
the blue and green curves represent the coverage of CO and the fraction
of the $1 \times 1$ phase respectively.
The coverage of CO and the fraction of the $1 \times 1$ phase at the center
of the laser spot are marked by red crosses.
Parameters used: domain length $L=800$, space discretization
$\Delta x=0.25$ with periodic boundary conditions.
The local temperature field around the laser spot is approximated
by a Gaussian profile ($\sigma=2$), with a maximum temperature
increase $\Delta T=5$K at the laser spot center $x=700$.
Reaction conditions: $T=541$K, $p_{\rm CO}=4.68\times10^{-5}$
mbar, $p_{\rm O_2}=1.33\times10^{-4}$ mbar.
The unit dimensionless time and length correspond to real values
of 1 s and $\sim5.6$ $\mu$m respectively.
The unit dimensionless speed thus corresponds to a real speed of
$\sim5.6$ $\mu$m/s. }\label{enhance_isotropic}
\end{figure}

For simplicity, we first consider the case where the
diffusion of CO in our one-dimensional domain is
isotropic (i.e. constant $D_u$).
With this assumption, the coverage profile of the adsorbates in
a {\it static} observation frame eventually becomes periodic in both
space and time, when a laser beam is
periodically scanned across the domain at a fixed speed $c$.
This oscillatory solution (in the static observing frame)
corresponds to a {\it steady state} solution in the co-moving
frame (i.e. of Eqs.~(\ref{cotravel_u})-(\ref{cotravel_w})).
A Newton-Raphson iteration combined with pseudoarclength
continuation is used on
Eqs.~(\ref{cotravel_u})-(\ref{cotravel_w}) to compute and continue
the corresponding steady coverage profiles as the laser dragging
speed, $c$, is varied (that is also the speed of the resulting
dragged pulse).
The CO$_2$ production rate is then averaged over the entire domain for
each steady coverage profile and plotted against $c$, as
shown in Fig.~\ref{enhance_isotropic}.
The reported enhancement is in comparison to the reaction rate at the
quenched state (without laser illumination).
Note that the steady profiles of adsorbate coverages computed
through pseudoarclength continuation may be unstable (the
dashed line in Fig.~\ref{enhance_isotropic}), which can lead to an
estimate of the average CO$_2$ production rate very different from direct
simulations (through time integration) at fixed $c$.
We will discuss this in detail at the end of this section.

In Fig.~\ref{enhance_isotropic}, when the laser dragging speed is
increased from zero to approximately 23 (at point 2), the steady
profiles of the adsorbate coverage retain a pulse-like appearance.
Such pulse-like profiles gradually start to ``lag behind"
the laser spot, as the laser dragging speed increases, which leads
to a monotonic increase in the overall CO$_2$ production rate to
20\% compared to the quenched state.
Following the solution branch from point 2 to point 4 in the Figure, the
pulse-like structure gradually dissolves and  a ``plateau" in the
CO coverage profile starts to develop  between the center of the
laser spot and the lagging pulse-like structure.
The ``pulse" now
appears to consist of two distinct fronts - a leading and a trailing one.
After point 4, the plateau also starts to shrink and finally
disappears around $c=24$.
Such a process occurs in a very narrow range of $c$ (approximately
between 23 to 24) and causes a very sharp increase in the overall CO$_2$
production rate enhancement (up to 70\%) followed by an immediate
sharp decrease (down to approximately 5\%), which is apparently
associated with the development and disappearance of a wide plateau
in the CO coverage profile.
When the laser scanning speed is increased above 24, the rate
enhancement quickly drops to zero.

\begin{figure}
\centering
\includegraphics[width=0.9\columnwidth]{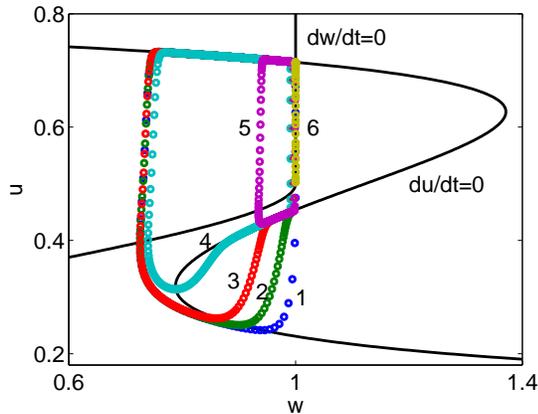}
\caption{(Color online) Phase portraits (circles) for steady coverage profiles at
different $c$ values (points 1 to 6 as in
Fig.~\ref{enhance_isotropic}), see text.}\label{dragging_profile}
\end{figure}

The emergence of the pulse-like structure and the plateau in CO
coverage profile is thus seen to be intimately related to the rate
enhancement.
In order to better understand the nature of this relation, we
explore the connection between the steady spatial coverage
profiles and the dynamics of the KEE ODE model, as shown in
Fig.~\ref{dragging_profile}.
It has been argued ~\cite{Bar94} that, since the oxygen adsorption is much faster
than CO desorption, the temporal variation of oxygen coverage can
be adiabatically eliminated for simplification, leading to a reduced
two-variable model (for $u$ and $w$).
The corresponding nullclines of this two-dimensional model are plotted in
Fig.~\ref{dragging_profile} (solid lines).
Representative dragged pulse coverage profiles in the co-moving
frame (e.g. points 1 to 6 in Fig.~\ref{enhance_isotropic})
correspond to limit cycles (periodic trajectories) of the ODE
system (in $z$) obtained by setting the left side of
Eqs.~(\ref{cotravel_u})-(\ref{cotravel_w}) to zero.
These $u$ and $w$ profiles are superposed on the simplified
two-equation model nullclines in Fig.~\ref{dragging_profile} (circles).

The trajectories corresponding to points 1 to 3 ``reach down" to the lower branch of
the $du/dt=0$ nullcline, confirming the pulse-like nature
of the pattern that results from laser-dragging.
The CO plateau structure, clearly seen in the spatial coverage
profiles corresponding to points 3 to 5, are associated with the profile
remaining {\em close to the middle branch} of the $du/dt=0$ nullcline.
This ``overlap" with the middle branch shrinks as
the laser scanning speed $c$ is increased.
In dynamical systems terminology, this plateay structure is thus
associated with a ``canard" \cite{Moehlis02}.
For point 6, the trajectory does not attain the middle branch of the
$du/dt=0$ nullcline; it aligns with the $dw/dt=0$
nullcline.

Based on the discussion above, the relationship between the
enhancement in CO$_2$ production rate and the laser dragging speed
$c$ , as shown in Fig.~\ref{enhance_isotropic}, can be partially
rationalized as follows:
When $c$ is increased from zero up to values in the neighborhood of point 2 ($\sim
23.3$), the laser spot locally excites the catalytic surface,
creates and ``drags along" a reacting pulse (e.g. point 1).
Such a pulse remains attached to the laser spot; it starts becoming more
spatially extended as $c$ is increased, leading to a monotonic
increase in production rate (appreciable adsorbed CO and oxygen coexisting
over longer spatial extents).
When $c$ becomes too fast for the laser actuation to excite the local
catalytic surface, the brief temperature increase effected by the
passage of the laser spot only leads to a small local variation in the coverage profile
(e.g. point 6).
For values of $c$ between the two limiting cases above (e.g.
between points 2 and 5 in Fig.~\ref{enhance_isotropic}), the steady
coverage profiles computed from pseudoarclength continuation become
unstable.
In this case, direct simulations must be used to find what the
stable long-term behavior is, and thus establish a
connection between $c$ and the corresponding rate enhancement
(see below).

\begin{figure}
\centering
\includegraphics[width=0.8\columnwidth]{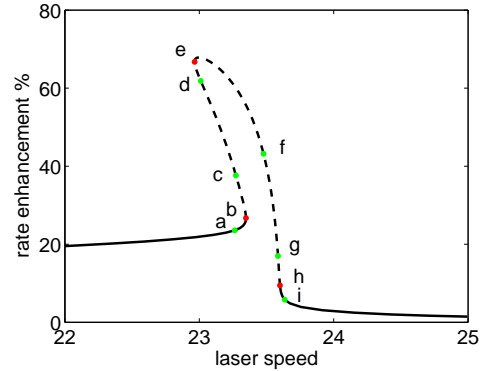}\\\vspace{0.4in}
\includegraphics[width=0.75\columnwidth]{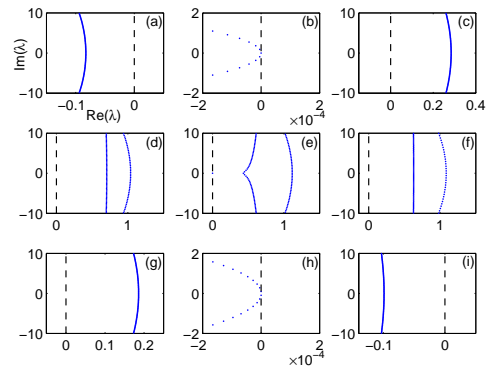}
\caption{(Color online) A blow-up picture of the unstable solution branch (dashed
line) in Fig.~\ref{enhance_isotropic}.
Bifurcation points are marked by red dots.
Following the solution branch from left to right,
representative eigenvalue spectra (marked by dots)
corresponding to the labelled points are plotted in the insets
below. }\label{enhance_unstable}
\end{figure}

A blow-up of the unstable solution branch in
Fig.~\ref{enhance_isotropic} is shown in
Fig.~\ref{enhance_unstable} along with the
corresponding eigenvalue spectra of the system
linearization in the neighborhood of the bifurcation points.
These linearization spectra have been computed in a finite
but long domain with periodic boundary conditions; computed
eigenvalues are marked by dots, and the density of these dots
is suggestive of continuous spectrum components for the infinite domain problem.

As we move from point $a$ to point $c$, a ``parabola" of eigenvalues crosses the
imaginary axis from left to right.
For our large but finite periodic problem, the leading eigenvalue (the eigenvalue(s) with
largest real part on this parabola) is a single real one
(as in inset $(b)$).
The initiation of the crossing of such an ``eigenvalue parabola" corresponds to a
Saddle-Node (SN) bifurcation followed by a large number of subsequent Hopf
bifurcations.
As $c$ varies from point $c$ to point $d$, the leading spectrum component
appears to split in two.
At point $e$ a single real eigenvalue (whose path started close to the
parabolic spectrum component farthest to the left from the imaginary axis)
crosses the imaginary axis,
corresponding to another SN bifurcation.
From points $f$ to $g$, the split spectrum components appear merge again into
a single component, which then crosses the imaginary axis from right to left
(see point $h$).
The leading eigenvalues of this new parabolic spectrum component
(the last ones to cross to stability) are now a conjugate
pair (in contrast to point $e$).
The crossing of such a component of the spectrum appears in our
finite domain case as (from point $g$ to point
$i$) as a chain of Hopf bifurcations.
Continuous eigenvalue spectrum crossing has been shown to be
associated with the emergence of ``chemical turbulence" for CO
oxidation in an excitable medium ~\cite{pwork7}.

\begin{figure}
\centering (a)\hspace{1.4in}(b)\\
\includegraphics[width=0.48\columnwidth, clip]{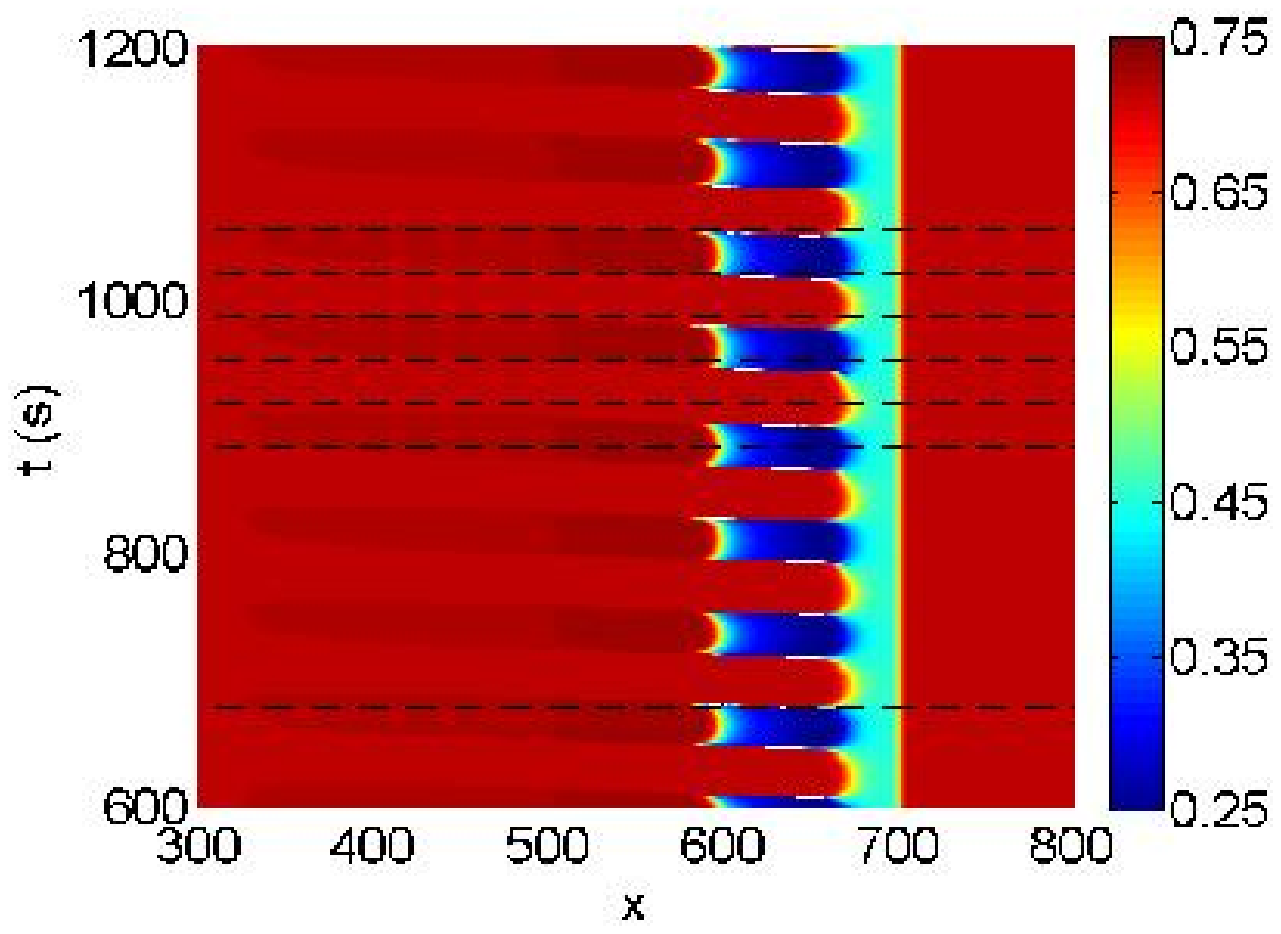}
\includegraphics[width=0.47\columnwidth]{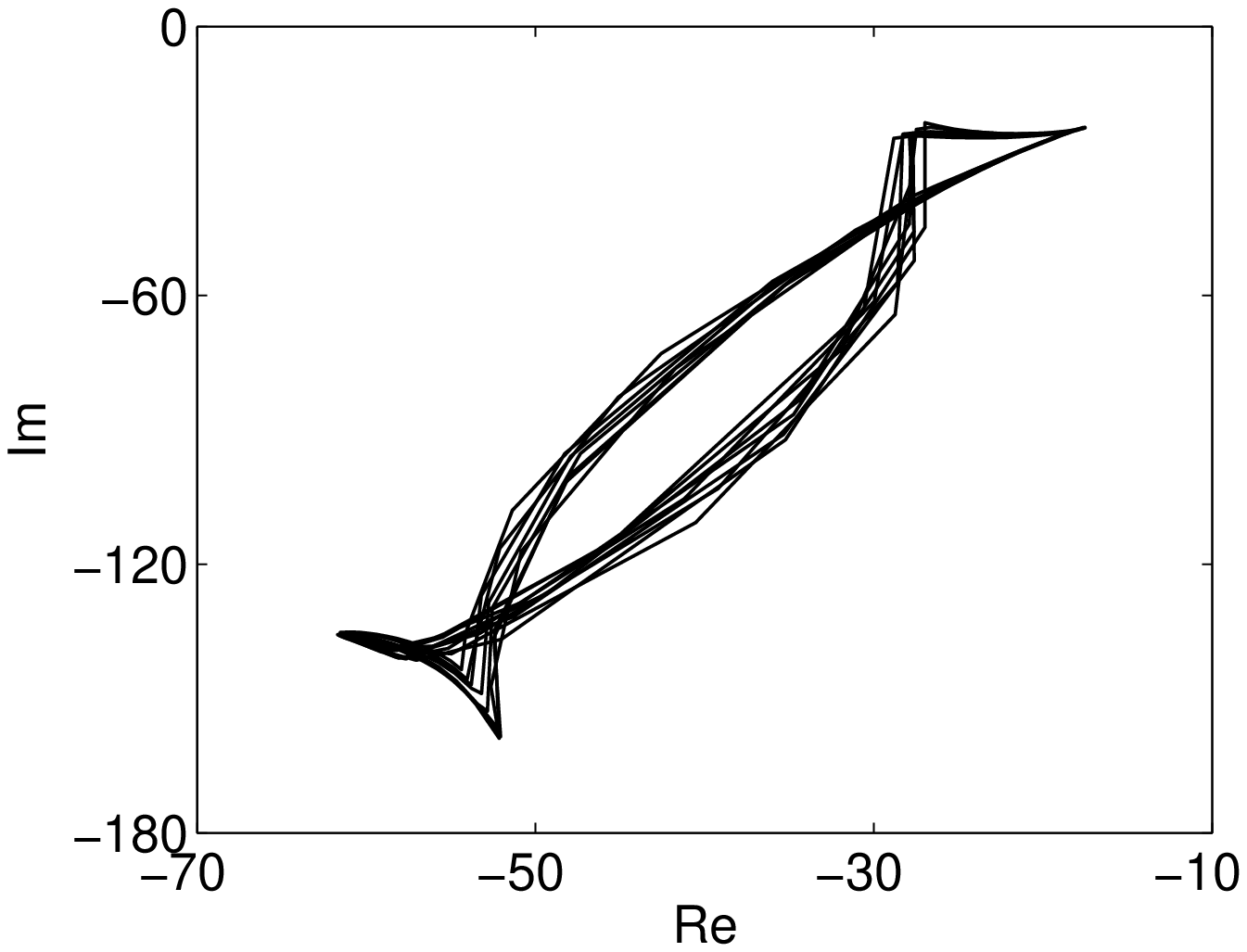}\\
(c)\hspace{1.4in}(d)\\
\includegraphics[width=0.48\columnwidth, clip]{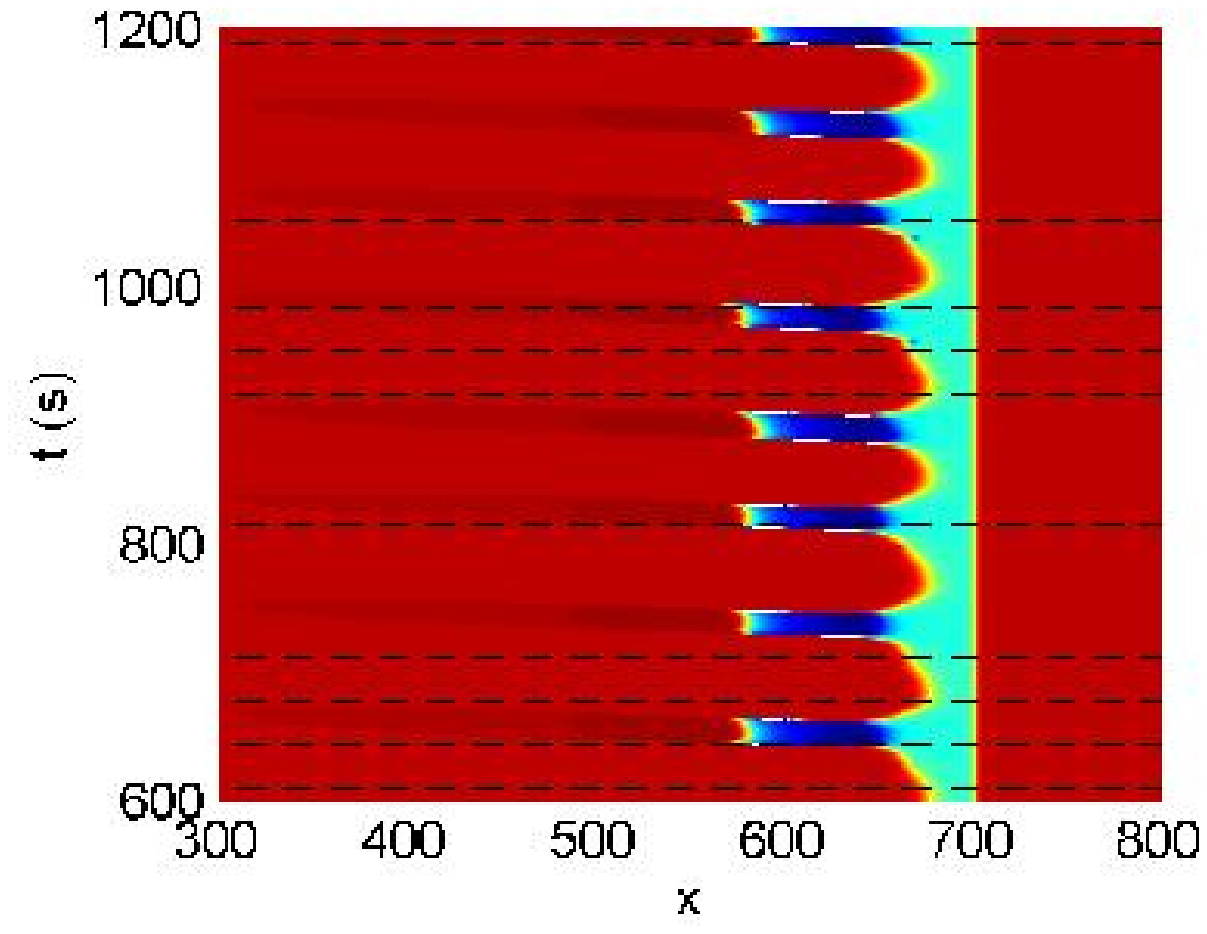}
\includegraphics[width=0.47\columnwidth]{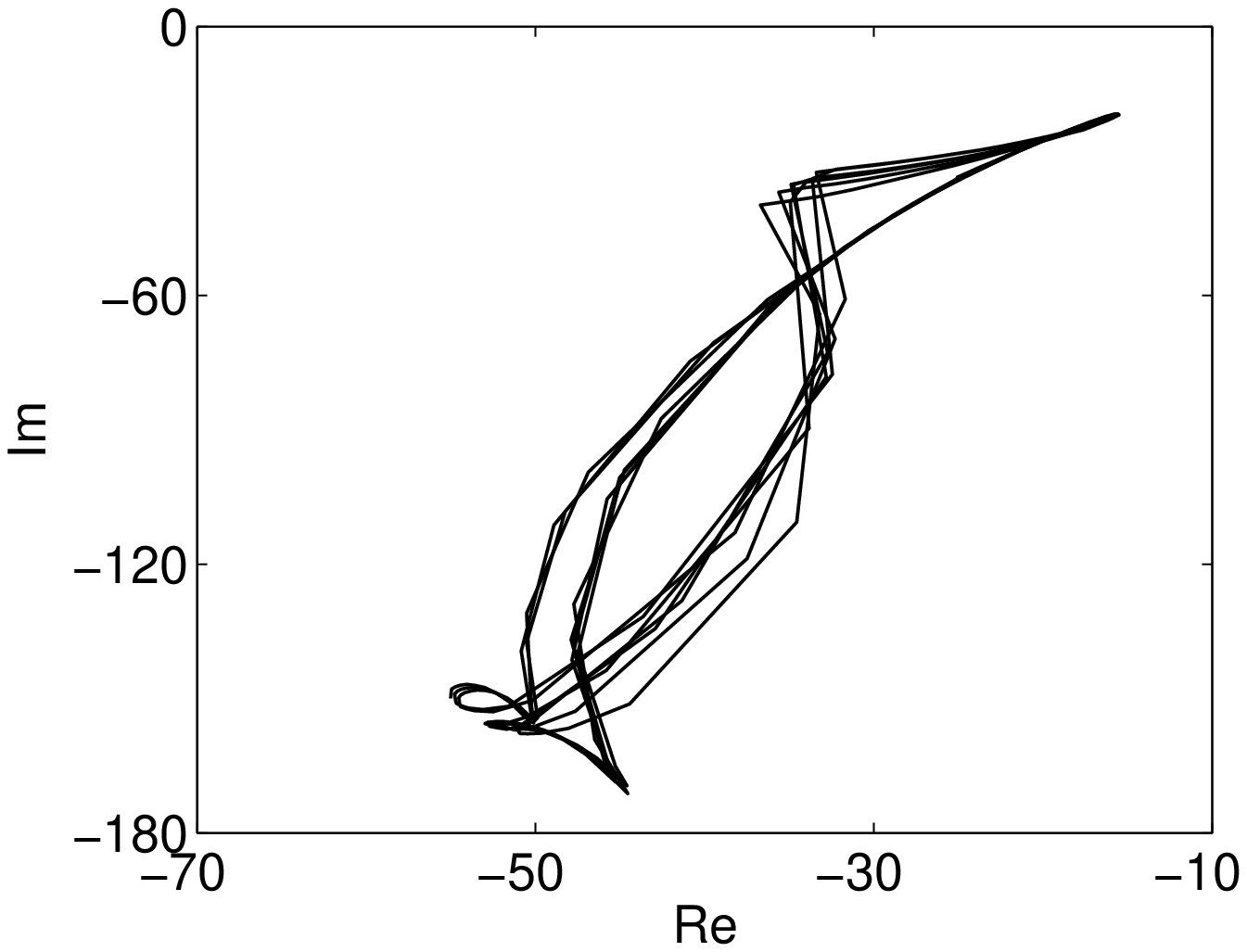}
\caption{(Color online) Space-time plots and phase portraits of CO coverage dynamics
for different laser scanning speeds in the unstable regime.
(a) and (c), space-time plots of CO coverage for $c=23.5$ and
$23.6$ respectively.
(b) and (d), the corresponding phase portraits of the first fast
Fourier transform component of the CO spatial coverage profile.
When $c$ is increased the system spends less time on
average at a state with a long reacting tail; thus
the overall CO$_2$ production rate is less in (c)
as compared with (a).
The time interval between dashed lines in (a) and (c) corresponds to
$L/c$, that is, one ``rotation" of the laser beam along the periodic domain.
\label{dragging_isotropic}}
\end{figure}

For values of $c$ for which the dragged pulse solutions are
unstable, direct simulation is used to establish the nature of
the observed dynamics (and to find the corresponding average reaction
rate).
Phase diagrams of the long-term dynamics for such simulations appear
quasiperiodic or mildly chaotic, and do not repeat
exactly (as shown in Fig.~\ref{dragging_isotropic}).
For sufficiently large domains (as was the case here), the
approximate period of the oscillations is not affected by the
domain length.
In this regime, as $c$ is increased, the system spends less time
on average in states with a long, reacting pulse-like tail, which
leads to a decrease in the overall (space-time averaged) CO$_2$ production rate.
This is consistent with the (spatiotemporally) averaged reaction rate
monotonically decreasing as
$c$ varies from points $e$ through $h$ in
Fig.~\ref{enhance_unstable}.

\begin{figure}
\centering
\includegraphics[width=0.9\columnwidth]{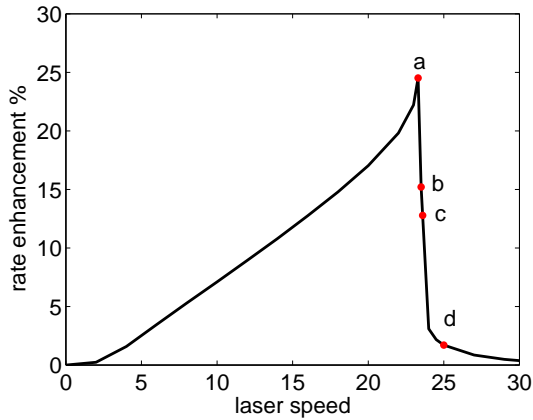}
\caption{Averaged rate enhancement in CO$_2$ production as a
function of laser scanning speed for anisotropic CO diffusion.
For a fixed laser scanning speed, the CO$_2$ production rate is
computed based on the spatiotemporal coverage profile, averaged
over the entire domain and a sufficiently long time interval.
$D_{uf}$ and $D_{us}$ are estimated from experimental
data~\cite{Oertzen94}; $D_{uf}/D_{us}=3$.}
\label{enhance_anisotropic}
\end{figure}

\begin{figure}
\begin{center}
\centering (a)\hspace{1.4in}(b)\\
\includegraphics[width=0.48\columnwidth, clip]{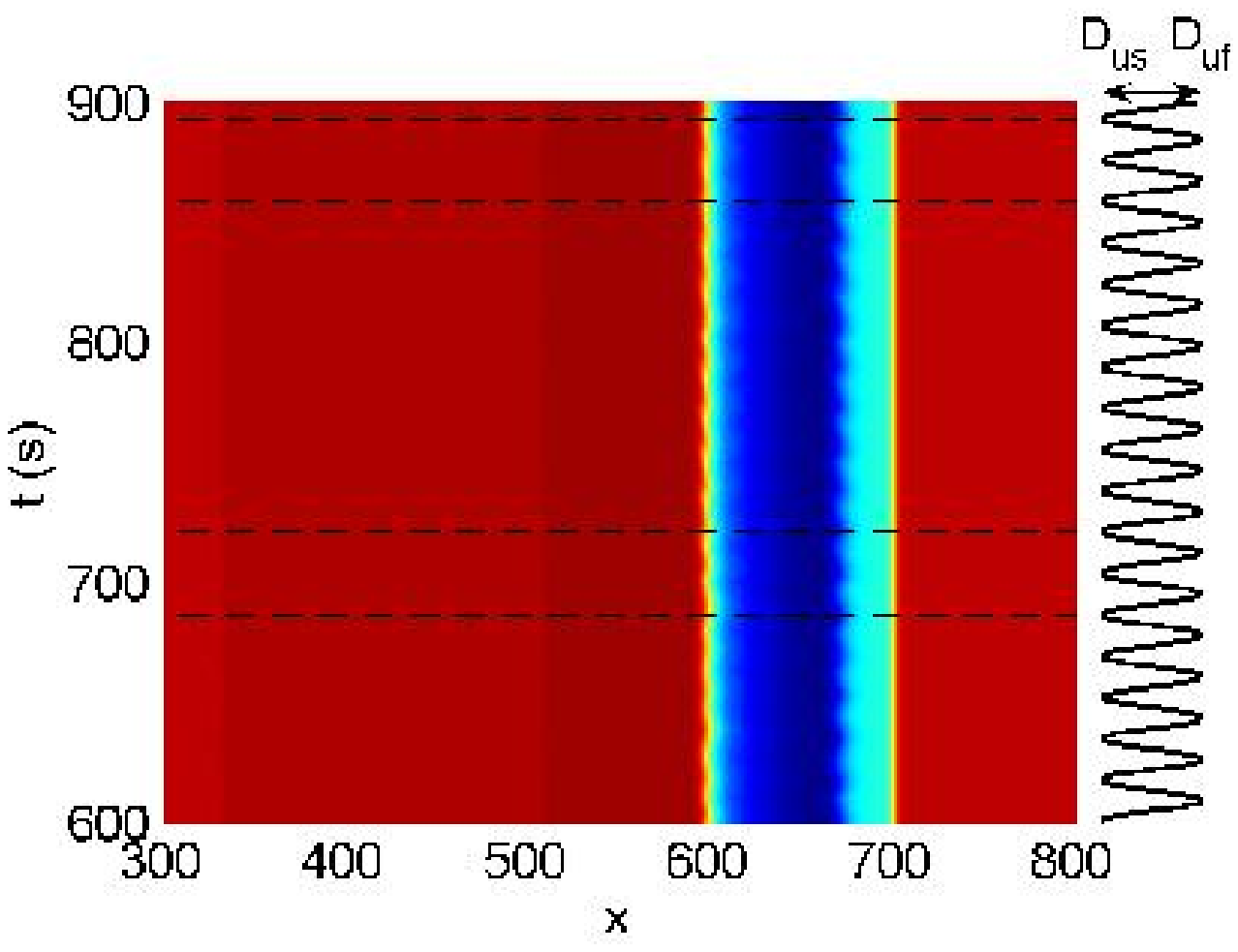}
\includegraphics[width=0.48\columnwidth, clip]{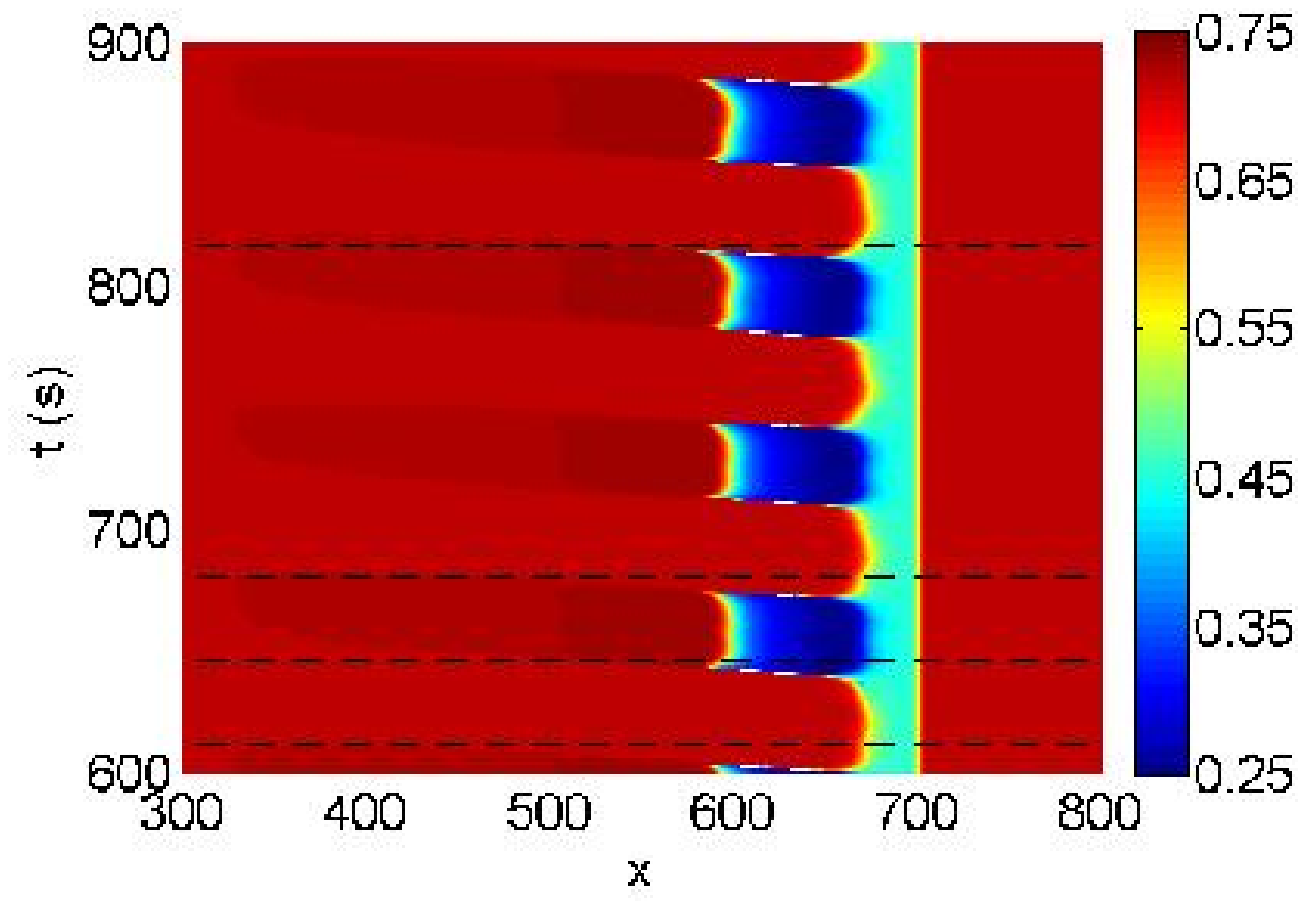}
(c)\hspace{1.4in}(d)\\
\includegraphics[width=0.48\columnwidth, clip]{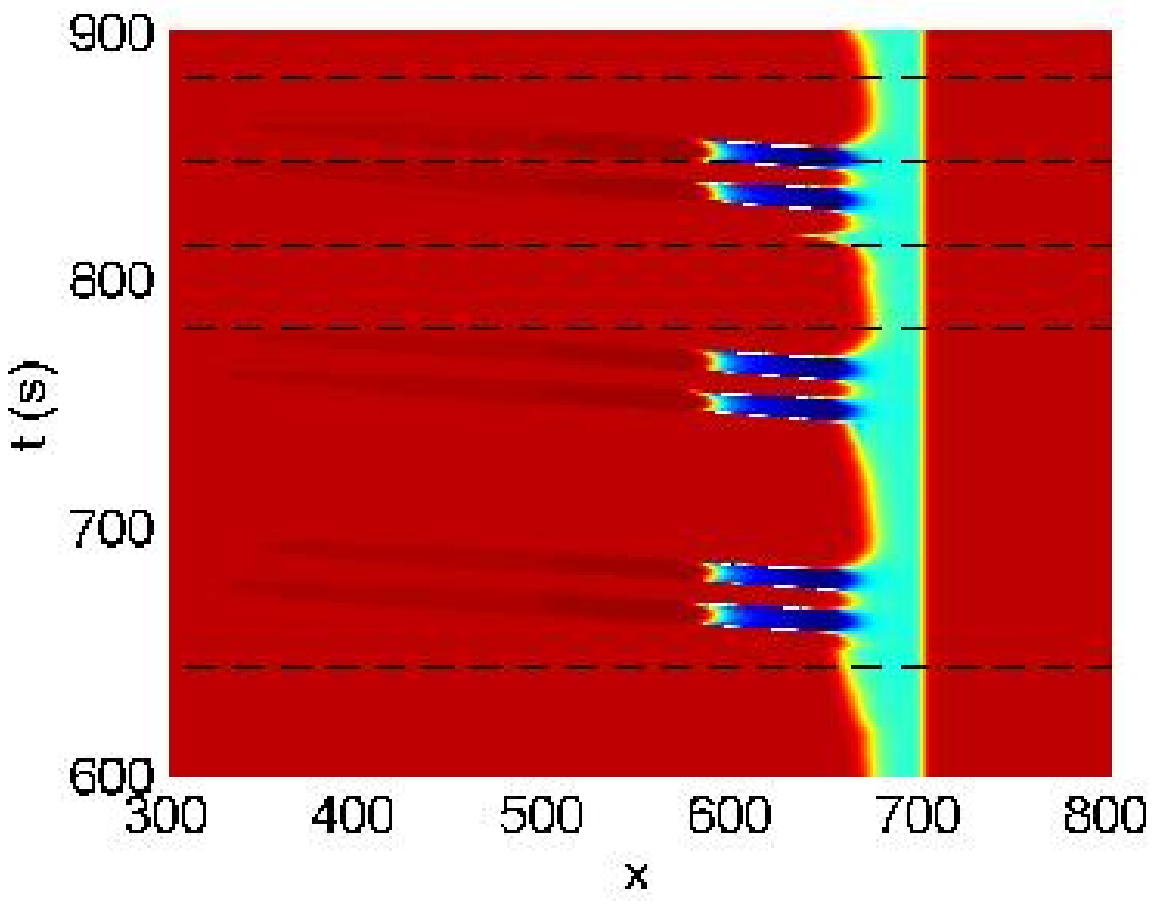}
\includegraphics[width=0.48\columnwidth, clip]{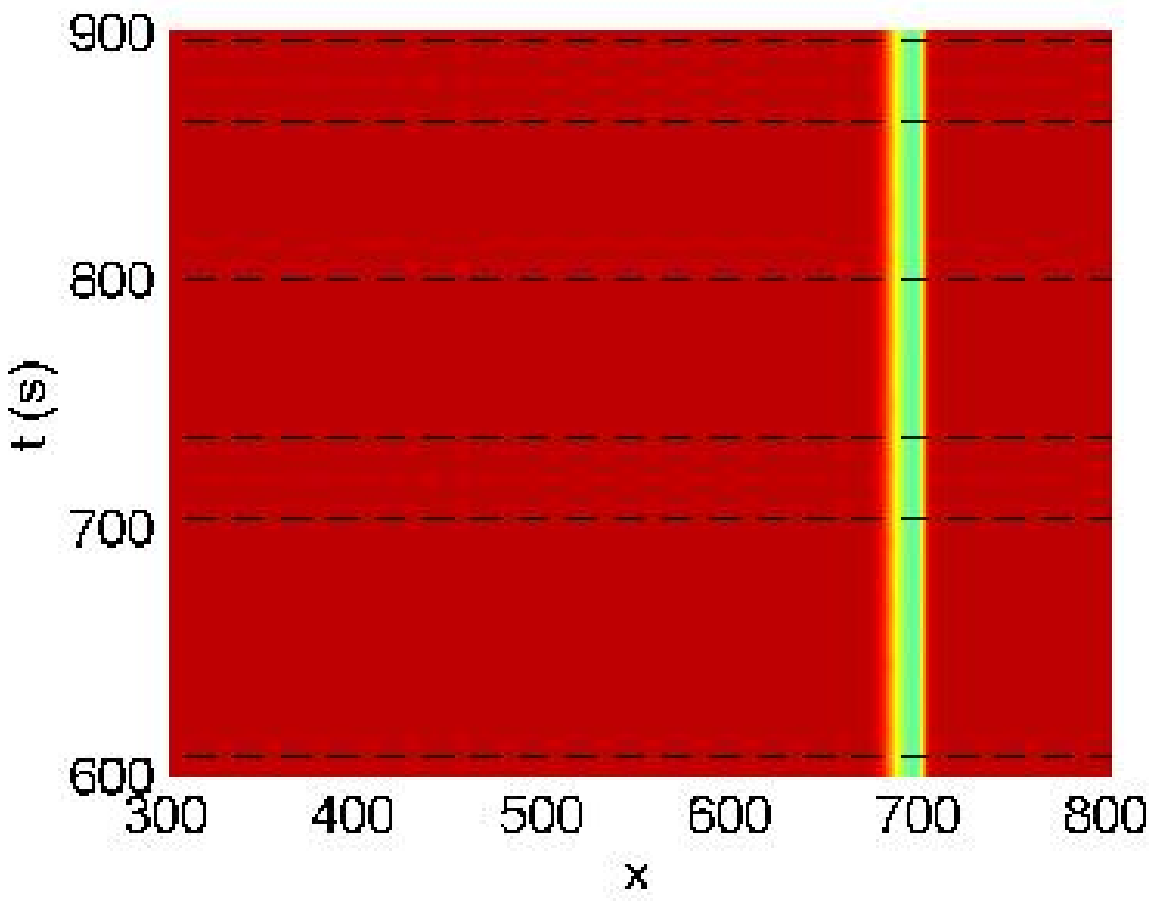}
\end{center}
\caption{(Color online) Phase portraits for anisotropic CO diffusion at different
laser scanning speeds.
The local CO diffusion coefficient at $x=700$ is plotted on the
right of plot (a).
(a) $c=23.3$, (b) $c=23.5$, (c) $c=23.6$, (d)
$c=25$.}\label{dragging_anisotropic}
\end{figure}

{\bf Anisotropic CO Diffusion.}
To incorporate the effect of diffusion anisotropy for the adsorbed
CO on Pt(110), we approximate the anisotropic diffusion
coefficient with a spatially periodic function,
$D_u\equiv(D_{uf}+D_{us})/2+(D_{uf}-D_{us})/2\times sin(4\pi
x/L)$, where $D_{uf}$ and $D_{us}$ correspond to the coefficients
of CO  diffusion on Pt(110) in the fast ([1$\bar1$0]) and in the slow ([001]) directions
respectively.
\begin{figure} [t]
\centering (a)\hspace{1.4in}(b)\\
\includegraphics[width=0.49\columnwidth, clip]{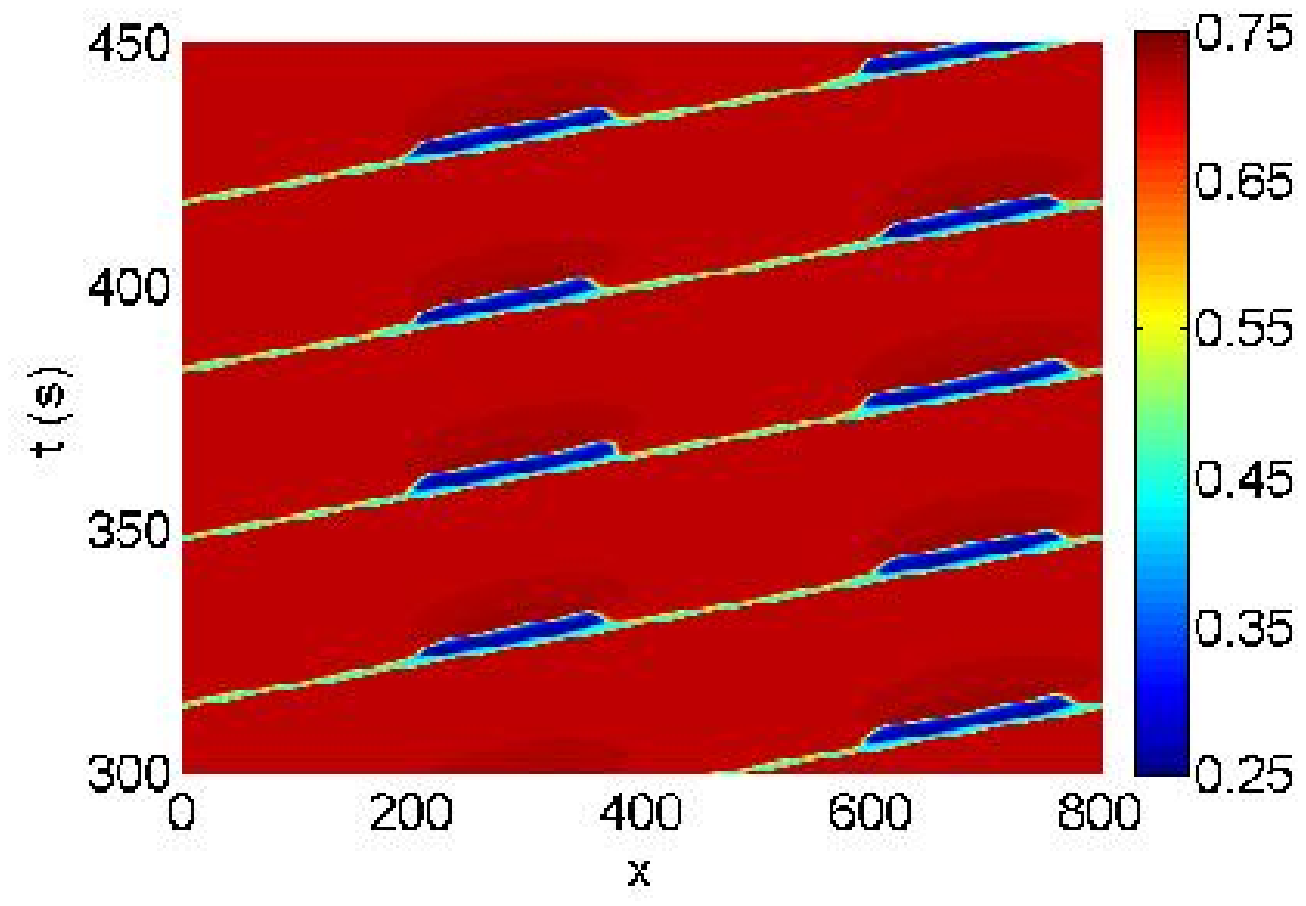}
\includegraphics[width=0.47\columnwidth]{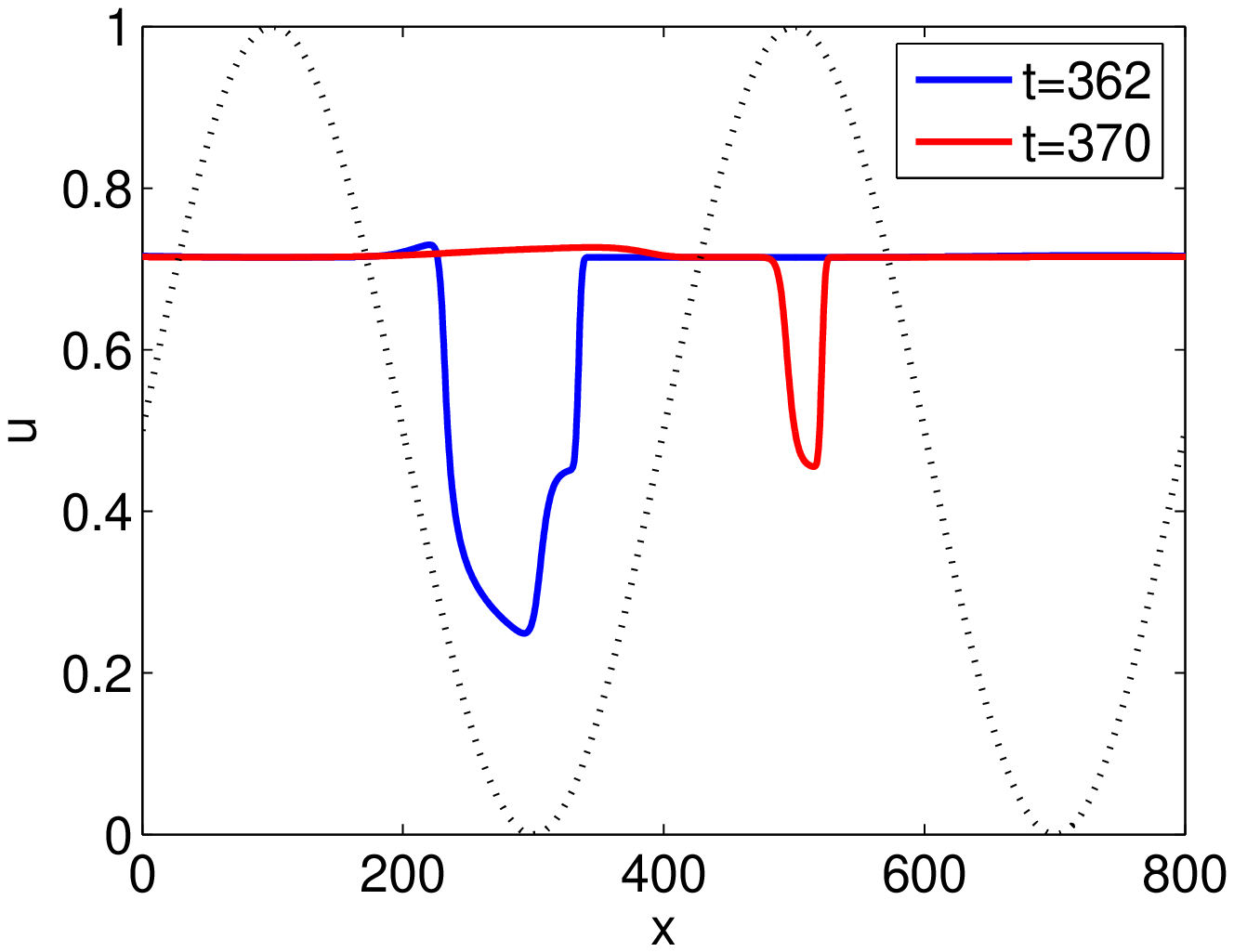}
\caption{(Color online) Laser scanning in the presence of strong anisotropic CO diffusion.
(a) space-time plot of CO coverage in a frame co-taveling with the laser spot;
(b) representative instantaneous snapshots of the CO coverage profile.
The anisotropic diffusion of CO is approximated by a periodic
diffusion coefficient with a period of $L/2$ (dotted line in (b)),
$D_u\equiv(D_{uf}+D_{us})/2+(D_{uf}-D_{us})/2\times sin(4\pi
x/L)$, where $D_{uf}$ and $D_{us}$ correspond to the diffusion coefficient
values along the fast and slow directions on Pt(110) respectively.
%
%
$D_{uf}/D_{us}=3$, $c=23.3$.}\label{strong_anisotropy}
\end{figure}
The interaction between laser scanning and strong anisotropic CO
diffusion is illustrated in Fig.~\ref{strong_anisotropy}, which
shows that the reacting tails only exist in regions where the
local CO diffusion is slow.
To visually enhance this phenomenon, and also to show how different the
instantaneous pulse profiles can be as the laser rotates on the surface
Fig.~\ref{strong_anisotropy} employs
diffusion coefficient values six times as large as the ones reported in~\cite{Oertzen94}
(but with the correct anisotropy ratio).
Resetting $D_{uf}$ and $D_{us}$ to their experimentally reported
values~\cite{Oertzen94}, time integrations of
Eqs.~(\ref{cotravel_u}) to~(\ref{cotravel_w}) show periodic or
apparently quasiperiodic transient behavior
(Fig.~\ref{dragging_anisotropic}).
In both cases we average the reaction rate
over both the domain length and a long enough time interval to compute
the corresponding average enhancement in CO$_2$ production rate shown
in Fig.~\ref{enhance_anisotropic}.
When $c$ is varied between 0 and approximately 23, the dynamics
in the co-moving frame
converge to a stable oscillatory state with a period of $L/(2c)$.
During the oscillations, the length of the reacting tail only
slightly varies as the local CO diffusion coefficient modulates (not shown).
%
%
Similar behavior has been observed for larger $c$ (i.e. $>24$),
where the width of local coverage profile around the laser spot
slightly modulates.
Compared with the isotropic case, the averaged rate enhancement
shows the same qualitative behavior for the
two limiting cases: for smaller $c$ the rate
enhancement gradually increases as $c$ is increased; for large $c$ the
rate enhancement suddenly decreases as $c$ is increased.
When $c$ is in the intermediate range $23\sim24$ (associated with
instability in the isotropic case), the system can show distinct
quasiperiodic transient behavior: for $c$ close to the lower
boundary of this interval the system exhibits oscillations similar to the
corresponding isotropic case (Fig.~\ref{dragging_anisotropic}(b));
for $c$ close to the upper boundary, ``skipping" oscillations can
be observed, as shown in Fig.~\ref{dragging_anisotropic}(c).
By comparing the position of reacting tails in
Fig.~\ref{dragging_anisotropic}(c) with the corresponding spatial
profile of $D_u$, we find that the reacting tails occur only along
the slow diffusion directions (when the local $D_u$ is small).
This suggests that the local periodic change in CO diffusion
coefficient as the laser spot moves along a ring can have a strong impact
on the local dynamics; it can, for example temporarily eliminate a reacting
tail and then reestablish it (see the small time gap in
Fig.~\ref{dragging_anisotropic}(c) when the reacting tail disappears).
As in the isotropic case, the averaged rate enhancement in
this range of $c$ monotonically decreases as $c$ is increased.

\section{Experiments}

The simplest motion of a temperature heterogeneity on the
platinum surface would be a linear one, on an infinitely extended
catalyst.
This cannot be realized in the experiment.
The easiest consistent
approach is to use a protocol where the laser spot moves on a
circle with large circumference.
This is experimentally feasible
and makes a comparison with theory possible.
Circular movement was tentatively explored
in a previous study~\cite{twists}; however, the
resulting surface dynamics and changes in reaction rate were not
analyzed in detail.
Here, we show new experimental results,
focussing on reaction enhancement and the influence of the laser
spot on the reaction patterns; we also compare to the
theoretical results presented in the first part of the paper.


Experiments are performed using a Pt(110) single crystal with
10~mm diameter located in an ultra high vacuum (UHV) chamber.
For
accurate control of the reaction parameters (temperature, CO and
oxygen partial pressures), the chamber is equipped with a computer
controlled gas dosing system, that allows for adjusting the
partial pressures of carbon monoxide and oxygen.
The platinum
sample is heated from the back with a halogen lamp.
For sample
preparation and characterization, the chamber is equipped with
Ar-ion sputtering and Low-Energy Electron Diffraction (LEED).
A
tube ($\sim$38~mm in diameter) with a cone-like ending ($\sim$4 mm
in diameter) is placed about 2~mm away from the sample surface.
It connects the reaction chamber with a differentially pumped
Quadrupole Mass Spectrometer (QMS) used for analysis of the
reaction products.
This way, a large fraction of the carbon dioxide produced
during the reaction can be detected, providing an estimate of the
overall surface reaction rate.
The carbon monoxide signal detected
by the QMS was stored on a computer.

\begin{figure}
\centering
\includegraphics[width=0.8\columnwidth]{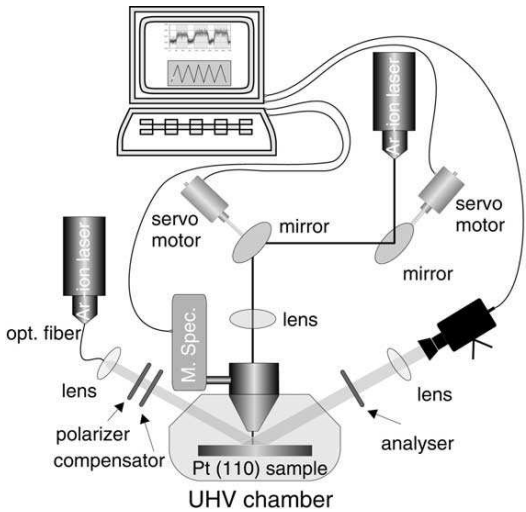}
\caption{Schematic view of the experimental setup, not to scale.
See the text for details. Reproduced
from~\cite{twists}.}\label{enhancement_setup}
\end{figure}

Concentration patterns of CO and oxygen on the sample surface are
imaged at a rate of 25 images per second using Ellipsomicroscopy
for Surface Imaging (EMSI)~\cite{rotermund-surfsci386-1997}.
They are recorded with a video CCD camera.
The background of the video signal is
subtracted and the signal is contrast-enhanced in real time using a
downstream Argus~20 image processing unit from Hamamatsu and
stored on a video recorder.
The surface reaction can be locally
manipulated by a Argon ion laser beam which is focussed onto the platinum
crystal.
The spot size of the laser is approximately 75~$\mu$m in diameter.
Two computer-controlled mirrors allow for controlled motion of the laser spot on
the sample surface.
Special motion protocols (like linear or
circular trajectories of the laser spot) can be programmed using
LabView.
The partial pressures of oxygen and CO are
also controlled through this software.
The experimental setup is sketched
in Fig.~\ref{enhancement_setup}.

%

\begin{figure}[t]
\centering
\includegraphics[width=0.9\columnwidth, clip]{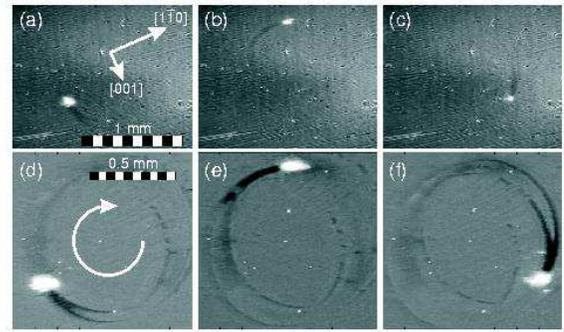}
\caption{Circular motion of the laser spot across the platinum
catalyst. (a)-(c) raw data, (d)-(f) images after additional background
subtraction, contrast enhancement and cropping.
The time
difference between the snapshots is 0.96~s.
The crystal
orientation is indicated in (a).
Diffusion along the [1$\bar1$0] axis is faster (by an estimated
factor of 2-3) than in the [001] direction.
The laser spot is
visible as a bright white area in the images.
Due to the laser motion, a
``tail" develops behind the spot.
The reaction parameters are $P_{\rm Laser}$=640~mW, $p_{\rm
O_2}$=$3.00\times10^{-4}$~mbar, $p_{\rm
CO}$=$8.43\times10^{-5}$~mbar, and $T=513$~K.}\label{Fig_Exp1}
\end{figure}

At the beginning of each experiment, the laser beam is focussed to
a fixed position on the sample surface for several minutes, until
the temperature of the crystal equilibrates.
This is necessary
because the laser spot not only produces a strongly located
temperature increase but additionally slightly increases the mean crystal
temperature (by a few K).
The laser beam is then scanned across the
sample surface with a constant speed for 40 seconds, following a
fixed circular path with a circumference of approximately 2.7~mm.
This is illustrated in Fig.~\ref{Fig_Exp1}.
The scanning of the
laser beam is stopped for a minute every time before a different
scanning speed for the laser beam is applied; at each speed
several measurements are performed and their results in terms of
reaction rate increase are averaged.

%
\begin{figure}
\centering
\includegraphics[width=0.9\columnwidth]{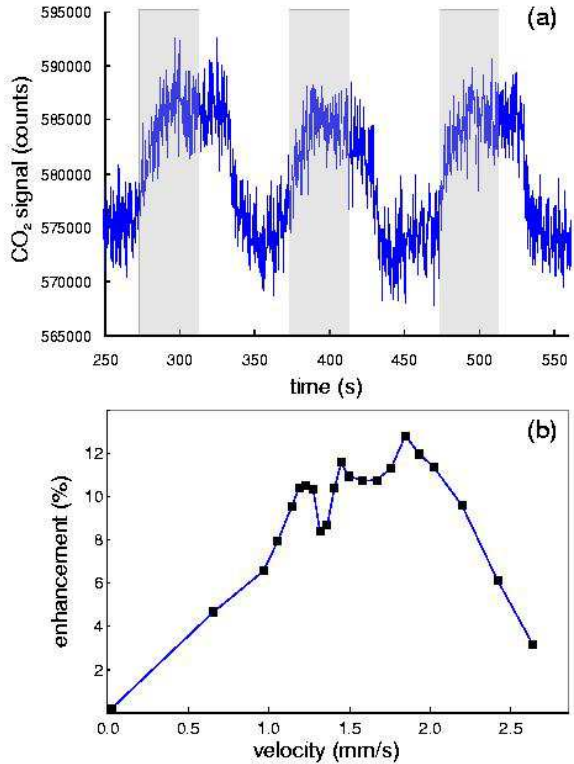}
\caption{Enhancement of CO$_2$ production.
(a) QMS signal; the
laser spot is only moving during the shaded time intervals with a velocity
of 0.91~mm/s.
(b) Average enhancement for different velocities of the
laser spot.
The reaction parameters are $P_{\rm Laser}$=640~mW, $p_{\rm
O_2}$=$3.00\times10^{-4}$~mbar, $p_{\rm
CO}$=$8.60\times10^{-5}$~mbar, and $T=512.5$~K.}\label{Fig_Exp2}
\end{figure}

In EMSI images, dark surface area is associated with more
reactive, predominantly oxygen covered surface.
When the laser
spot moves, a dark tail can be observed: the laser spot
removes some adsorbed CO in its path.
Since CO
poisons the catalytic surface, its removal increases the catalytic
activity.
In addition to the one-dimensional theoretical observations in the first
part of this article, we see here the full two-dimensional structure of
the laser induced reactive surface area.
At a certain distance
from the laser spot, the tail divides, and a ``swallowtail" composed of two narrow
``tongues" can be seen.
Figure~\ref{Fig_Exp2}(a) shows the CO$_2$ signal measured by the
QMS.
During the periods of laser spot motion
(shaded regions) the CO$_2$ signal is increased.
The delay between stopping the laser spot motion
and associated decrease of the CO$_2$ production
rate is due to the slow time scale of the differential pumping of
the QMS.
%
%

We measure the increase of the
carbon dioxide signal as a function of the laser spot velocity.
The results are shown in Fig.~\ref{Fig_Exp2}(b).
We observe that the average reaction rate increases for velocity values
up to about 2.8~mm/s.
For higher velocities, the enhancement falls precipitously
until the laser spot moves so fast that no enhancement
can be measured.
When all other reaction parameters are kept fixed, there exists a
certain velocity of the laser spot at which the enhancement is at
it's maximum.
We call this velocity value the ``optimal velocity".
The experimental curve agrees qualitatively with theory: A slow
increase of the reaction rate is followed by a fast decrease after
the optimal velocity has been exceeded.
This has been observed in our computations both for isotropic and
for anisotropic diffusion coefficients; in the case of anisotropic
diffusion the dragged pulse profile varies along the laser path
because the diffusion coefficient also varies.
The variation in the length of the pulse tail in the
three snapshots of Fig.~\ref{Fig_Exp1} is the result of the
varying orientation of laser motion with respect to the
crystal axes (and the associated CO diffusion coefficient variation).

\begin{figure}
\centering
\includegraphics[width=0.9\columnwidth]{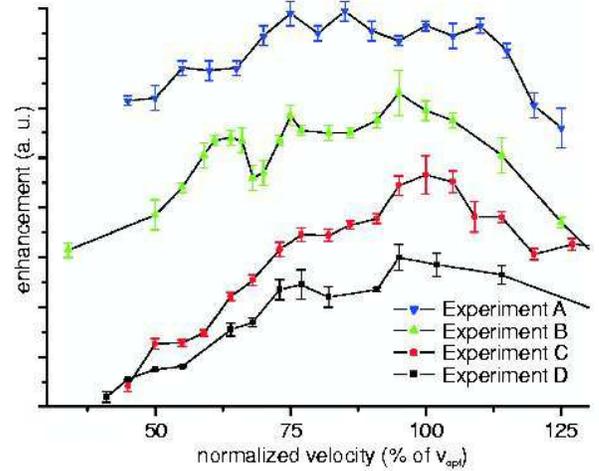}
\caption{Reaction rate enhancement at different experimental conditions:
Each enhancement curves is normalized with respect to its
optimum velocity and arbitrarily shifted with respect to the y axis for
better comparison.
The reaction conditions were $P_{\rm Laser}$=640~mW, $p_{\rm
O_2}$=$3.00\times10^{-4}$~mbar,
$p_{\rm CO}$=$8.60\times10^{-5}$~mbar, and $T$=512.5~K for
experiments A and B,
and $p_{\rm CO}$=$8.43\times10^{-5}$~mbar, and $T$=513~K for
experiments C and D.}\label{Fig_Exp3}
\end{figure}
Figure~\ref{Fig_Exp3} shows the results of experiments at
different reaction parameters, normalized with respect to the
optimal velocity of each experiment.
The qualitative behavior of the reaction
enhancement appears similar in all cases shown.
The experimental
curves exhibit a clear maximum in the overall rate enhancement as
the laser scanning speed is varied, in agreement
with the computational results shown above.
It is interesting to observe certain secondary reaction
extrema ``on the way" to the overall maximum as the
laser spot velocity increases, e.g. at
$v$~=~1.35~mm/s in Fig.~\ref{Fig_Exp2}(b).
These were not observed in our one-dimensional modeling, and
they most probably should be
attributed to the two-dimensional nature of the system.
%

\begin{figure}
\centering
\includegraphics[width=0.9\columnwidth, clip]{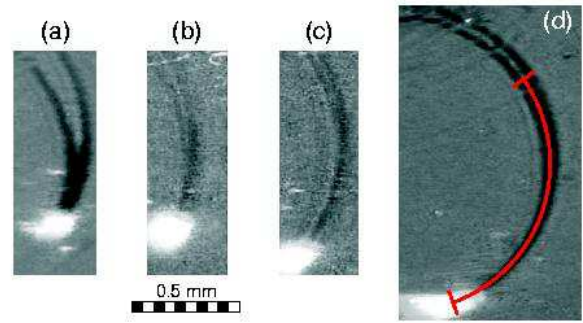}
\caption{Length of the dark reactive tail at different velocities.
(a) $v$=0.94~mm/s, $l$=0.23~mm, (b) $v$=1.2~mm/s, $l$=0.38~mm, (c)
$v$=1.8~mm/s, $l$=0.53~mm, (d) $v$=2.6~mm/s, $l$=0.79~mm,
parameters like in Fig.~\ref{Fig_Exp1}.}\label{Fig_Exp4}
\end{figure}


As the laser moves on the catalytic surface, it locally shifts
the system into an
 excitable state.
This results in the
development of a reactive ``tail" behind the laser spot.
The length
and the width of this tail vary with the spot velocity as
clearly seen in Fig.~\ref{Fig_Exp4}.
Qualitatively, the variation of the average tail length with
the laser dragging speed mimicks the variation of the
reaction rate enhancement both in modeling and in
experiments: the tail initially becomes longer
and then precipitously disappears as the speed is increased.
However, due to the development of two
``tongues" in the back of the tail, a clear definition of the tail
length in the experiment is somewhat difficult.
We (rather arbitrarily) chose the distance between the laser spot and the
point where the tail visibly splits as a measure of the tail length.
With
increasing laser speed the tail becomes longer and narrower.
In Fig.~\ref{Fig_Exp4}~(d) the length of the tail has
increased further, yet the reaction enhancement is almost
negligible.
This appears counterintuitive, since in the theoretical part
of this article we saw a clear correlation between tail length and
reaction rate enhancement.
However, the one-dimensional modelling ignores the width (and overall
two-dimensional structure) of the tail; this makes direct
comparison between the two-dimensional experiment and the
one-dimensional theory difficult.
We expect to report on a more quantitative comparison,
based on full two-dimensional simulations, in
a future publication.
%
%
For very
high laser speeds in the experiment -as well as in the 1D model-
the reactive tail completely vanishes, since the thermal energy deposited per time
and surface area in the catalytic surface no longer provides a
sufficient local temperature increase to excite the system.

\section{Summary and conclusions}

Following our initial experimental study~\cite{twists} on
improving catalytic surface activity through time-dependent
operating policies, we performed here a systematic exploration of
the effect laser spot motion on the average reaction rate of CO oxidation on
a Pt(110) surface.
Quantitative relations between overall rate enhancement and the
laser scanning speed, for both isotropic and
anisotropic CO diffusion, have been obtained through numerical
bifurcation analysis and direct simulations.
A characteristic local maximum in the rate enhancement
(gradual increase followed by precipitous drop) has been observed in
both cases.
The local maximum for isotropic CO diffusion was associated
with the development of a pulse instability caused by laser
dragging and apparently involving the dragged pulse continuous spectrum.
When anisotropic CO diffusion is included in the model, and for a wide range
of laser scanning speeds, the system exhibits comparable
{\it averaged} rate enhancement response; yet the anisotropy has a significant
qualitative effect on the detailed dynamics (including quasiperiodicity
and even mild spatiotemporal chaos).
The computational results are supported by experimental observations
showing a qualitatively comparable increase of averaged reaction rate with
laser scanning velocity.
Optimal velocities were obtained experimentally for a range of different
reaction conditions.
Experimental observations revealed an interesting two-dimensional
structure of the dragged pulse ``reactive tail"; this clearly suggests
that full two-dimensional modeling, taking into account diffusion
tranverse to the pulse motion, is required for a more quantitative
comparison of theory with experiments.
We are currently working towards such a fully two-dimensional model study.
The laser motion explored in this paper was, in some sense, the
simplest spatiotemporal one: effectively one-dimensional with
constant speed.
Spatiotemporally more complex motions, such as those presented in
\cite{twists}, possibly also containing a stochastic component
\cite{Wolff03}, should provide a rich field of study.

{\bf Acknowledgements}  This work was partially supported by the
National Science Foundation and by a Guggenheim Fellowship
(I.G.K.). The experiments were performed while two of the authors (C.P., H.-H.R.)
were at the Fritz Haber Institute of the Max Planck Society (Physical Chemistry) in
Berlin.


\end{document}